\documentclass[useAMS,usenatbib,usegraphicx]{mn2e}

\newcommand{\beq}	{\begin{equation}}
\newcommand{\eeq}	{\end{equation}}
\newcommand{\beqa}{\begin{eqnarray}}
\newcommand{\eeqa}{\end{eqnarray}}

\def\simlt{\lower.5ex\hbox{$\; \buildrel < \over \sim \;$}}
\def\simgt{\lower.5ex\hbox{$\; \buildrel > \over \sim \;$}}
\def\la{\simlt}
\def\ga{\simgt}
 
\font\tenbi=cmmib10 \newfam\bifam  \textfont\bifam=\tenbi

\font\tenbr=cmbx10
\newfam\brfam  \textfont\brfam=\tenbr

\font\squinttenbi=cmbx10 at 9pt
\scriptfont\brfam=\squinttenbi

\def\vecnabla{
              \setbox1=\hbox{$\bigtriangledown$}
                           \raise.45ex\hbox{$\bigtriangledown$\hskip-.97\wd1
                           $\bigtriangledown$\hskip-.97\wd1
                           $\bigtriangledown$\hskip-.97\wd1}
                           \raise.47ex\hbox{$\bigtriangledown$}}

\def\rsun{\ifmmode {\rm R}_{\mathord\odot}\else $R_{\mathord\odot}$\fi}
\def\msun{\ifmmode {\rm M}_{\mathord\odot}\else $M_{\mathord\odot}$\fi}
\def\lsun{\ifmmode {\rm L}_{\mathord\odot}\else $L_{\mathord\odot}$\fi}

\newcommand{\kms}	{{\rm km}\, {\rm s}^{-1}}

\def\tmb{\ifmmode {T_{\rm mb}^{13}(x,y,v)}\else $T_{\rm mb}^{13}(x,y,v)$\fi}

\newcommand{\xco}       {X$_{\rm CO}$~}
\newcommand{\xc}        {X$_{\rm CI}$~}
\newcommand{\htwo}      {{H$_{2}$}~}
\newcommand{\wco}       {$W_{\rm CO}$~}

\newcommand{\xunits}     {cm$^{-2}$K$^{-1}$km$^{-1}$s}
\newcommand{\nhtwo}        { $N_{{\rm H}_2}$}

\title[The $X_{\rm CI}$-Factor]{An Alternative Accurate Tracer of Molecular Clouds: The ``$X_{\rm CI}$-Factor''}
\author[Offner et al.]
{Stella S.~R.~Offner$^1$\thanks{Hubble Fellow, stella.offner@yale.edu}, Thomas G.~Bisbas$^2$, Tom A.~Bell$^3$, and Serena Viti$^2$\\
$^1$Department of Astronomy, Yale University, New Haven, CT 06511 \\
$^2$Department of Physics and Astronomy, University College London, Gower Street, London WC1E 6B \\
$^3$Centro de Astrobiolog\'ia (CSIC-INTA), Carretera de Ajalvir, km 4, 28850 Madrid, Spain
}

\begin{document}

\pagerange{\pageref{firstpage}--\pageref{lastpage}} \pubyear{2013}

\maketitle

\label{firstpage}

\begin{abstract}
We explore the utility of CI as an alternative high-fidelity gas mass tracer for galactic molecular clouds. We evaluate the ``X$_{\rm CI}$-factor'' for the 609 $\mu$m carbon line, the analog of the CO ``X-factor'', which is the ratio of the \htwo  column density to the  integrated $^{12}$CO(1-0) line intensity. 
We use {\sc 3d-pdr} 
to post-process hydrodynamic simulations of turbulent, star-forming clouds.
We compare the emission of CI and CO for model clouds irradiated by 1 and 10 times the average background and demonstrate that CI is a comparable or superior tracer of the molecular gas distribution for column densities up to $6 \times 10^{23}$ cm$^{-2}$.  Our results hold for both reduced and full chemical networks. 
For our fiducial Galactic cloud we derive an average \xco of $3.0\times 10^{20}$\xunits~  and \xc of $1.1\times 10^{21}$ \xunits.
\end{abstract}

\section{Introduction}

Within the local universe star formation occurs solely in molecular gas. 
Probing the distribution and dynamics of molecular gas is essential to understand the process by which gas converts into stars.
While H$_2$ is the most abundant molecule within molecular clouds by four orders of magnitude, it has no permanent dipole moment and, thus, no transitions that are excited under typical cold cloud conditions. The next most abundant molecule, CO, is uniformly adopted by the star formation community to probe molecular gas (see review by \citealt{bolatto13} and references therein).

However, the relationship between \htwo and other molecules is not trivial. 
In the observational domain where there is finite spatial resolution, limited sensitivity, and significant uncertainties, chemical subtleties are intractable. Consequently, a constant value known as the ``X-factor'' is widely used to connect CO emission and H$_2$ gas mass.
The X-factor is defined as the ratio of the total \htwo column density, \nhtwo, to the integrated $^{12}$CO ($J=1\rightarrow0$) line emission, \wco:
\beq
X_{\rm co} = \frac{N_{\rm H_2}}{W_{\rm CO}}.
\eeq
For Milky Way molecular clouds \xco$\sim 2 \times 10^{20}$ \xunits~ (e.g., Table 1 in \citealt{bolatto13}). 
Given that the 
cloud environment 
 is observationally difficult to constrain, gas properties inferred using the standard \xco 
 are likely prone to large errors. 


Other molecules, such as HCN, and gas proxies, such as dust emission and extinction, provide alternative means of constraint albeit with additional challenges and uncertainties (e.g., \citealt{gao04,bolatto11}). One promising tracer is atomic Carbon,  CI via the $^3P_1 \rightarrow ^3P_0$ transition, which appears to be correlated with both CO and \htwo abundances, at least in low-density environments \citep{papadopoulos04,bell07}. This CI transition corresponds to 492.16 GHz (609\,$\mu$m) in contrast to the first rotational transition of CO, which is 115.27 GHz (2.6 mm).  To date, a variety of observational studies have found strong similarity between CI and CO isotopologues \citep{plume94,plume99,ikeda99,ikeda02,kulesa05,shimajiri13}. The correspondence between the CI and $^{13}$CO emission is even more pronounced than for $^{12}$CO \citep{ikeda99,plume99,ikeda02,shimajiri13}.
Historically, widespread CI emission associated with molecular clouds received little attention since it was assumed that it traced a thin photodissociation region (PDR) layer on the surface of the cloud  \citep{tielens85,plume94,hollenbach99,plume99}. However, if CI exists mainly in a surface layer, this is a curious correlation, since $^{13}$CO traces denser gas on smaller scales than $^{12}$CO, which more quickly becomes optically thick.  

A variety of explanations have been proposed to explain the ubiquity of CI, including ``clumpy'' clouds  (e.g., \citealt{spaans97,cubick08}) and nonequilibrium chemistry (e.g., \citealt{deboisanger91,xie95}).
\citet{papadopoulos04} was the first to challenge the view of CI as a surface tracer.
Some combination of factors may be at play, but in essence, the observed emission can be explained only if CI and CO are more concomitant than past calculations suggest. 

The view of CI as a spatially limited ``surface'' tracer has its roots in simple one-dimensional PDR models. 
In this Letter,  we reevaluate this classical picture. We explore the utility of CI as a molecular gas tracer through comparison with CO, by examining  emission on a point-by-point basis in fully three-dimensional simulated clouds, which are meant to represent typical Milky Way clouds. We extend previous studies to consider higher number density environments, a full chemical network, and 3D turbulent cloud conditions. \S 2 reviews our hydrodynamic, chemical, and post-processing methods. In \S 3 we present our analysis comparing CO and CI, and we discuss the implications in \S 4. 



\begin{figure}
\vspace{-0.1in}
\includegraphics[width=8.0cm]{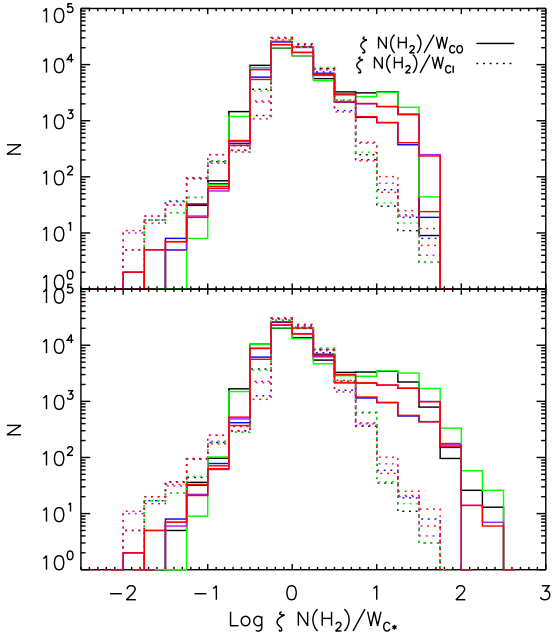} 
\caption{Distribution of CO(solid) and CI(dotted) X-factors for 6 orthogonal views for Rm6\_1.0\_12\_1a.  The distributions are normalized to their median value. The bottom panel includes all emission; the top only includes channels with $T>$0.1\,K. \label{wcodist} }
\end{figure}

\begin{figure}
\includegraphics[width=8.5cm]{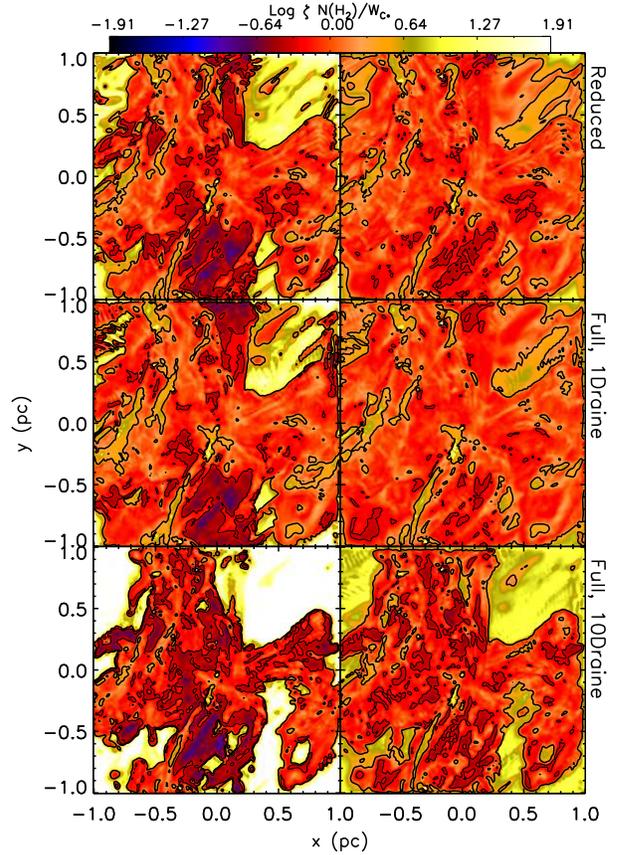}
\vspace{-0.25in}
\caption{Distribution of Log $\zeta$\xco (left) and Log $\zeta $\xc (right) for one view through the simulation for run Rm6\_1.0\_12\_1a (top), Rm6\_1.0\_12\_1f (middle) and Rm6\_1.0\_12\_10f (bottom).  The X-factors are normalized to their median value, $1/\zeta$. Contours show a factor of 2 above (thick) and below (thin) the median.  \label{wcodistmap} }
\end{figure}

\begin{figure}
\vspace{-0.1in}
\includegraphics[width=8.5cm]{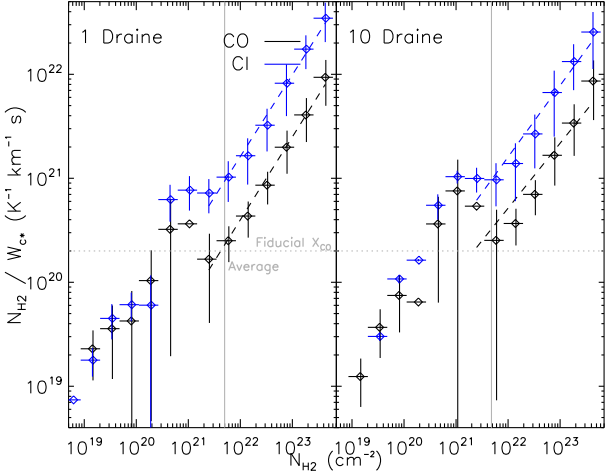}
\vspace{-0.15in}
\caption{CI (blue) and CO (black) mean X-factors as a function of \htwo column density for one view  of Rm6\_1.0\_12\_1f (left) and  Rm6\_1.0\_12\_10f (right). The means are obtained from the Figure \ref{wcodistmap} data by binning the pixels as a function of their column density. The vertical error bars indicate the standard deviation and the horizontal error bars indicate the bin size. The grey lines show the fiducial CO X-factor (horizontal dotted) and the mean column density (vertical solid). The dashed lines are least-squares best fits to the data. An online-only figure shows the mean $W_{\rm C*}$ as a function of $N_{\rm H_2}$.\label{xvals} }
\end{figure}

\begin{figure}
\vspace{-0.3in}
\includegraphics[width=8cm]{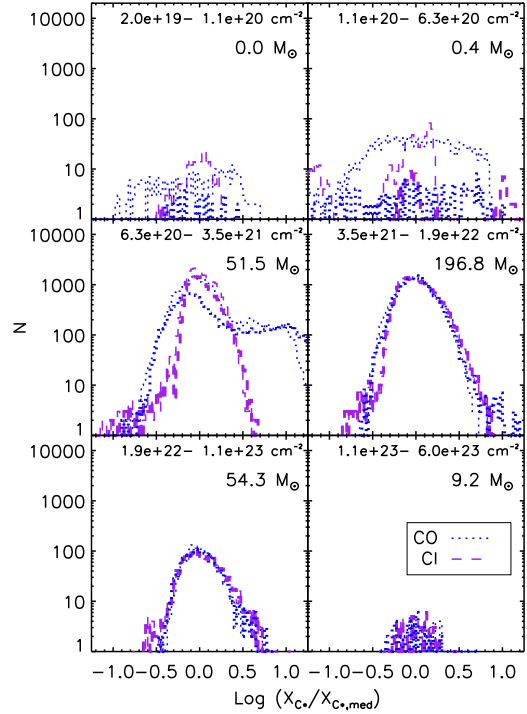}
\vspace{-0.1in}
\caption{Histogram of CI (dashed) and CO (dotted) X-factors as a function of \htwo column density for one view through run Rm6\_1.0\_12\_1f (thin) and  Rm6\_1.0\_12\_10f (thick). The \htwo column density range  and the total \htwo mass appear at the top. 
\label{wcodistbin} }
\vspace{-0.5in}
\end{figure}


\begin{table*}
\caption{{\sc 3d-pdr} Run Parameters and X-factor results. \label{simpdr}}
\renewcommand{\arraystretch}{0.5}
\tabcolsep=0.1cm
\begin{tabular}{@{}lccccc|cccccc}
\hline
Run ID$^a$  & FUV(G$_0$) & Network & $\bar n_{\rm H}$(cm$^{-3}$) & $\bar N_{\rm H_2}$(cm$^{-2}$) & $f$256$^3$  & $$$\bar{X_{\rm CO}}$ & $\alpha_{X{\rm CO}}$ & $X_{{\rm CO},0}$ & $\bar{X_{\rm C}}$ & $\alpha_{X{\rm CI}}$ & $X_{{\rm CI},0}$\\
\hline
Rm6\_1.0\_12\_1a & 1  & Reduced & 1080 & $3.0 \times 10^{21}$ & $1/12$ & $2.7\times10^{20}$  & 0.88 & $7.5\times 10^0$  & $6.8\times 10^{20}$ & 0.87 & $8.6 \times 10^1$  \\
Rm6\_1.0\_12\_1f & 1  & Full    & 1800 & $5.0 \times 10^{21}$ & $1/13$   & $3.0\times 10^{20}$ & 0.80 & $1.1\times10^{3}$ & $1.1\times 10^{21}$ & 0.79 & $8.4\times 10^{3}$\\
Rm6\_1.0\_12\_10 & 10 & Full    & 1800 & $4.8 \times 10^{21}$ & $1/13$   & $4.3\times 10^{20}$ & 0.62 & $9.4\times10^{6}$ & $1.2\times 10^{21}$ & 0.69 &$1.2\times 10^6$ \\ \hline

\end{tabular}
\medskip

$^a$The columns are: run ID,  magnitude of the external field, mean H-nucleus number density, mean \htwo column density, grid sampling used by {\sc 3d-pdr}, average CO X-factor, fitted \xco slope, fitted \xco intercept, average CI X-factor, fitted \xc slope, and fitted \xc intercept.  The fits only include pixels with $W>3\sigma$, where $\sigma_{\rm CO} = 0.15$ K$\kms$ and $\sigma_{\rm CI} = 0.2$\,K$\kms$ is the channel noise. \\
\end{table*}

\vspace{-0.2in}
\section{Numerical Methods} \label{methods}


We analyze a hydrodynamic simulation of a turbulent star-forming molecular cloud that was performed with the {\sc orion} adaptive mesh refinement (AMR) code; it is described in detail in \citet{offner13}, where it is the fiducial Rm6 run. We use the snapshot at one freefall time for our chemical post-processing. At this time,  17.6\% of the gas resides in ``stars'', which are modeled by sink particles \citep{krumholz04}. 
This simulation is meant to represent a $600\,\msun$ piece of a typical Milky Way cloud. Namely, the simulation velocity dispersion satisfies the observed linewidth-size relation \citep{mckee07}, the cloud is virialized \citep{tan06}, and the mean H$_2$ column densities are $\simeq 3-5 \times 10^{21}$ cm$^{-2}$  (Table \ref{simpdr}), which are similar to those of local star-forming clouds \citep{beaumont12}. Consequently, we expect the chemical distribution to reflect that of typical Milky Way star-forming regions and exhibit similar compositions of PDR and non-PDR gas.



We use {\sc 3D-PDR} \citep{bisbas12} to calculate the abundance distributions, heating and cooling functions, and the gas temperature at every point of the density distribution in the range $200\le n\le 10^5\,{\rm cm}^{-3}$. Full details of the code are given in \citet{bisbas12}, however in the simulations discussed here we use an updated technique to compute the rate of H$_2$ formation on grains using the treatment of \citet{caza02,caza04}. In addition, we determine the dust temperature following the treatment of \citet{Holl91}. We adopt two sets of the most recent UMIST2012 chemical database \citep{mcel13}; a ``reduced'' network consisting of 33 species and 330 reactions, and a ``full'' network of 215 species and $\sim$3000 reactions.  Heavier elements are included in the latter, such as Sulfur (S) and Iron (Fe), which affect the ionization fraction and thus alter the temperature deeper into the cloud \citep{jim11}, albeit at the cost of computational expense. Photodissociation of $H_2$ and CO is treated explicitly, including the effects of self-shielding.

The elemental abundances are set to values similar to those in local molecular clouds: $[{\rm He}]=1.0\times10^{-1}$,  $[{\rm C}]=1.41\times10^{-4}$,  $[{\rm O}]=3.16\times10^{-4}$,  $[{\rm Mg}]=5.1\times10^{-6}$, $[{\rm S}]=1.4\times10^{-6}$, $[{\rm Fe}]=3.6\times10^{-7}$. We consider isotropic radiation field strengths of $1\,{\rm Draine}$ and $10\,{\rm Draine}$,  where 1\,Draine is the interstellar radiation field defined by \citet{draine78}. 
In all {\sc 3d-pdr} models we use $12$ HEALPix rays \citep{gorski05}; a turbulent velocity of $1.5\,{\rm km}\,{\rm s}^{-1}$, similar to the hydrodynamic velocity dispersion; a cosmic-ray ionization rate of $5\times10^{-17}\,{\rm s}^{-1}$; and we evolve the chemistry for $10\,{\rm Myr}$ (see \citealt{bisbas12} for details). Table \ref{simpdr} lists the run properties.


We use the radiative transfer code {\sc radmc-3d}\footnote{{http://www.ita.uni-heidelberg.de/\~dullemond/software/radmc-3d}} to post-process the simulations and compute the CO and CI  emission. We adopt the non-LTE Large Velocity Gradient approach \citep{shetty11}, which computes the level populations given some density, abundance and temperature distribution. The total gas density and velocities are determined hydrodynamically by {\sc orion}, while the H, H$_2$, CO, and C abundances and gas temperature are determined by {\sc 3d-pdr}.  The H and H$_2$ abundances are required since they are the main collisional partners of CO and CI. We interpolate all the inputs to 256$^3$ resolution ($dx=$0.001 pc). For the excitation and collisional data, we adopt the Leiden database values \citep{schoier05}.

The output spectral data cubes have 256 channels spanning $\pm 10$ km s$^{-1}$ ($dv=0.08$ km s$^{-1}$). To account for velocity jumps between cells we use ``Doppler catching'' with $d_c=$0.025, which interpolates the velocities so changes between points do not exceed 0.025 times the local linewidth. We adopt a constant ``microturbulence'' of 0.1 km s$^{-1}$ to include sub-resolution line broadening caused by unresolved turbulence. Using {\sc radmc-3d}, we also compute the dust continuum emission assuming a standard ISM grain distribution and equal dust and gas temperature.

We use the Planck function in the Rayleigh-Jeans limit to convert from the {\sc radmc-3d} line intensity to brightness temperature.
The flux in a pixel is given by $W = \Sigma_i T_{B,i} dv$. 

\vspace{-0.5cm}
\section{Results}\label{results}

\subsection{X-factor Distributions}

If the line emission were perfectly correlated with the molecular column density then the distribution of X-factors would be a delta function. However, the correlation varies over the cloud as a function of varying temperature and abundance, so there is generally a wide range of X-factors. This has previously been investigated in some depth for CO in turbulent cloud simulations by \citet{shetty11}. 

Figure \ref{wcodist} shows the distribution of X factors for both CO and CI for 6 different lines of sight ($\pm x, \pm y, \pm z$) through the cloud. The narrower the distribution about a single value, the better the tracer reflects the underlying \htwo distribution. The top panel of Figure \ref{wcodist} shows the distribution of \xc-factors is more symmetric and narrower than the CO distribution. When the low-level emission is removed (bottom panel), the two distributions look more similar, although the CI is still slightly better. This indicates that the CI is definitely better for lower column density, fainter material, which is to be expected since that material is truly a PDR. CO underestimates the emission in such regions; however, this faint material may not be detectable for observations with low signal-to-noise.

Although six views are shown for each model, only a few different lines are immediately apparent. This occurs because most of the domain is optically thin such that views from opposite sides of the cloud show similar flux. Views along the three different cardinal directions produce fairly similar shaped distributions.

Figure \ref{wcodistmap} shows a map of the CO and CI X-factors. Perfect correspondence between column density and emission would yield a uniform map, however, this is clearly not the case. Substructure stands out in both cases but are more severe in the CO maps. This is a consequence of the column density increasing more strongly than the emission. The correlation between emission and column density should be worse in high-density regions (e.g., star-forming cores) where portions of the gas along the line of sight become optically thick and in low-$A_v$ regions where the CO begins to dissociate. Observationally, it is possible to use $^{13}$CO to correct for optical depth effects in high density regions, but this introduces new uncertainties (e.g., \citealt{heiderman10}).

\vspace{-0.2in}
\subsection{Column Density and FUV Field Dependence}

We find a relatively short range of column densities for which the X-factor is independent of density for both tracers. Figure \ref{xvals} shows that this occurs around the mean column density, where CO happens to be close to the fiducial \xco value.  In the higher column density regime, the X-factor increases strongly with column density. Above the mean value the data can be fit by the functional form
$X_{\rm C*} = X_0 N_{{\rm H}_2}^{\alpha_x}.$ 
Table \ref{simpdr} gives the best fit parameters according to a linear least squares fit. In Figure \ref{xvals}, we only include pixels with emission exceeding $3\sigma$, where we adopt typical noise values of $\sigma_{\rm CO}=0.15\, {\rm K\,}\kms$ \citep{pineda08} and $\sigma_{\rm CI}=0.2\, {\rm K}\, \kms$ \citep{shimajiri13}. By definition of the X-factor (Eq. 1), we expect $\alpha_x \simeq 0$ if the emission accurately traces column density. 
However, we find that both tracers have $\alpha_{x}\sim 0.6-0.8$. This is because the tracers become optically thick and are no longer sensitive to the column density, which is also indicated by the flattening of the line intensities at high column density (see Figure 5, available online). This was also found by \citet{shetty11} for CO.

Figure \ref{xvals} demonstrates that CI emission has less scatter for nearly all column densities, suggesting that a single X-factor may be a better approximation. While CO is expected to trace H$_2$ poorly in regions that have higher FUV flux, and hence are more photodissociated, $\bar X_{\rm CI}$ is nearly invariant.
 
For the sensitivity limits we adopt, the CI emission mass estimate is within 4\% and 7\% of the true \htwo mass for the 1 and 10 Draine fields, respectively. The CO emission overestimates the mass by 9\% and 47\%. These values are roughly proportional to the area of the yellow regions in Figure \ref{wcodistmap}. For this reason, $\bar X_{\rm CO}$, especially, depends on the sensitivity limit. Adopting $1.5 \sigma$ and $6\sigma$ gives $\bar X_{\rm CO}$ of $3.5 \times 10^{20}$ and $2.8 \times 10^{20}$ \xunits, while $\bar X_{\rm CI}$ remains unchanged.

Our fiducial $X_{\rm CO}$ is about $50\%$ higher than the standard value.  Our value of \xc is also somewhat higher than those calculated by \citet{bell07}, who found \xc=0.4-1.9$\times 10^{20}$ \xunits. However, these calculations assumed average galaxy-wide conditions and adopted higher FUV fluxes and significantly different metallicities as appropriate for extragalactic environments.


Figure \ref{wcodistbin} shows the distribution of X-factors as a function of different column density ranges. The CI distributions are comparable or narrower except for the lowest column density range, where CO dissociates. The \htwo gas mass in each range, as indicated on the plot, underscores that the lower column density gas contains significant mass and is not simply low-density PDR.
In comparing the different FUV strengths in Figure \ref{wcodistbin}, the higher external field increases the \xco dispersion at intermediate column densities. CI appears to be less sensitive to changes in the field.

\vspace{-0.2in}
\section{Discussion and Implications}\label{obs}

A variety of explanations have been proposed for the correspondence between CO and CI. We discuss these here in the context of our results.

The most commonly invoked explanation is that molecular clouds are thought to be very inhomogeneous, which facilitates the penetration of UV photons deep into the cloud.  Such``clumpy'' morphologies would also increase the surface area of the CI layer \citep{spaans97}.  
In run Rm6\_1.0\_12\_f 73\% of the mass sees FUV flux in the range 0.01-1.0 Draines and could be considered PDR.
\citet{offner13} compared the abundance distributions of CO and CI and found that there are a similar range of abundances down to FUV strengths of $\sim$0.01 Draine (see their Figure 10). Although CI abundance declines at higher $A_v$, there is still a range of high $A_v$-low FUV gas for which  $n_{\rm CI}/n_{\rm H}\sim 10^{-4}$.

Dynamical processes such as turbulent diffusion may serve to mix the gas and eliminate CI gradients in the cloud \citep{xie95}. 
This effect would be much larger in more violent environments such as in galaxy-merger driven star formation  and the Galactic Center \citep{tanaka11}, but this could contribute to CI/CO mixing in active regions such as the Carina molecular clouds  \citep{papadopoulos04}. Here, we include microturbulence to represent small scale unresolved turbulence but our results are computed in post-processing so that dynamical mixing does not contribute to the abundance distribution.

Non-equilibrium chemical processes can also create an elevated CI to CO ratio (e.g., \citealt{deboisanger91,xie95,oka01}). This effect will be most pronounced in young, lower density ($n<10^4$ cm$^{-3}$) clouds \citep{papadopoulos04}. Given that the \htwo formation time is a good proxy for chemical equilibrium \citep{hollenbach99}, we can assess the equilibrium state of our runs. The \htwo formation time is given by $t_{{\rm H_2}} \simeq 0.5 \times 10^6 T_{100}^{-1/2}\bar n_3$
where $T_{100}$ is the temperature normalized to 100 K, $\bar n_3$ is the average Hydrogen number density normalized to $10^3$cm$^{-3}$. 
Our results are analyzed at $10^7$ yr, which is $>>t_{\rm H_2}$, suggesting that non-equilibrium effects do not play a role here.

The prevalence of CI can also be enhanced by star formation, which contributes additional UV and cosmic ray flux. Both of these effects contribute to systematically higher CI to CO ratios (e.g., \citealt{papadopoulos04}). Since we do not take into account additional FUV flux from embedded star formation and we adopt a standard cosmic ray flux, this cannot explain our result. However, cosmic rays are the dominant source of heating at high $A_v$ (see Appendix in \citealt{offner13}) and may partially contribute to the widespread CI distribution we find.

An alternative interpretation to CI tracing a smooth surface layer is that the CI emission may be less temperature sensitive than CO.
\citet{plume99} found that the CI emission map was smoother than the $^{12}$CO emission. Recent observations of CI in the Orion A cloud also support this interpretation \citep{shimajiri13}. Furthermore, \citet{beaumont13} show that observed CO structure may not truely correspond to underlying density variations, which may contribute to the mismatch between H$_2$ and CO emission.

The sum of these effects, many of which are not included in our modeling, suggest that CI may be even more prevalent in local molecular clouds and a better tracer of molecular gas than we find here.
These results have important implications for the study of star formation and molecular clouds. Although many instruments have been designed with the detection of CO in mind, a number of instruments have frequency ranges and sensitives sufficient to detect widespread CI, including ALMA and SOFIA. We recommend future studies combine CO and CI data to obtain more accurate estimates of the molecular gas distribution. 


\vspace{-0.2in}
\section*{Acknowledgments}
The authors thank Jonathan Foster for helpful discussions and acknowledge support from NASA grant HF-51311.01 (S.S.R.O), STFC grant ST/J001511/1 (T.G.B), and a JAE-DOC research contract (T.A.B.). T.A.B also thanks the Spanish MINECO for funding support through grants AYA2009-07304 and CSD200900038. T.G.B. acknowledges the NORDITA program on Photo-Evaporation in Astrophysical Systems (June 2013).
The {\sc orion} and {\sc 3d-pdr} calculations were performed on the Trestles XSEDE cluster and DiRAC-II COSMOS supercomputer (ST/J005673/1, ST/H008586/1, ST/K00333X/1), respectively.

\bibliography{turb_pdr_pdf.bib}
\bibliographystyle{mn2e}

\end{document}